\documentclass[10pt,aps,prl,twocolumn,superscriptaddress]{revtex4}

\usepackage[utf8]{inputenc}
\usepackage{hyperref}
\usepackage{graphicx}
\usepackage{amsmath,amssymb}
\usepackage{color}
\usepackage{subfigure}
\usepackage{flushend}
\graphicspath{{Images/}}

\begin{document}

\title{Pulsatile microvascular blood flow imaging by short-time Fourier transform analysis of ultrafast laser holographic interferometry}

\author{L. Puyo}

\affiliation{
Centre National de la Recherche Scientifique (CNRS) UMR 7587, Institut Langevin. Fondation Pierre-Gilles de Gennes. Institut National de la Sant\'e et de la Recherche M\'edicale (INSERM) U 979, UPMC, Universit\'e Paris 7. Ecole Sup\'erieure de Physique et de Chimie Industrielles ESPCI ParisTech - 1 rue Jussieu. 75005 Paris. France
}

\author{I. Ferezou}

\affiliation{Unit of Neuroscience, Information and Complexity (UNIC), UPR CNRS 3293, France}

\author{A. Rancillac}

\affiliation{Brain Plasticity Unit, CNRS UMR 8249, ESPCI ParisTech, 10 Rue Vauquelin, 75005 Paris, France}

\author{M. Simonutti}

\author{M. Paques}

\author{J. A. Sahel}

\affiliation{Institut de la Vision, INSERM UMR-S 968. CNRS UMR 7210. UPMC. 17 rue Moreau, 75012 Paris. France}

\author{M. Fink}

\author{M. Atlan}

\affiliation{
Centre National de la Recherche Scientifique (CNRS) UMR 7587, Institut Langevin. Fondation Pierre-Gilles de Gennes. Institut National de la Sant\'e et de la Recherche M\'edicale (INSERM) U 979, UPMC, Universit\'e Paris 7. Ecole Sup\'erieure de Physique et de Chimie Industrielles ESPCI ParisTech - 1 rue Jussieu. 75005 Paris. France
}

\begin{abstract}
We report on wide-field imaging of pulsatile microvascular blood flow in the exposed cerebral cortex of a mouse by holographic interferometry. We recorded interferograms of laser light backscattered by the tissue, beating against an off-axis reference beam with a 50 kHz framerate camera. Videos of local Doppler contrasts were rendered numerically by Fresnel transformation and short-time Fourier transform analysis. This approach enabled instantaneous imaging of pulsatile blood flow contrasts in superficial blood vessels over 256 $\times$ 256 pixels with a spatial resolution of 10 microns and a temporal resolution of 20 ms.
\end{abstract}

\maketitle

Microvascular blood flow plays a central role in fundamental physiology and in the pathophysiology of many diseases, yet imaging of hemodynamic signals down to capillaries without contrast agent remains a challenging task. Functional magnetic resonance imaging~\cite{OgawaLee1990} and functional ultrasound~\cite{MaceMontaldoCohen2011, MaceMontaldoOsmanski2013} are capable of imaging hemodynamic parameters non-invasively through several millimeters of living tissue, with a spatial resolution down to capillary levels. Optical techniques can also measure blood flow or oxygenation contrasts within microvessels in tissue at even higher spatial and temporal resolutions, but they are limited to superficial vascularisation imaging. Functional optical coherence tomography~\cite{Wang2010, DziennisQinShi2015} can reach intermediate depths (tens to hundreds of microns), and photoacoustic techniques~\cite{MaslovZhang2008, YaoMaslov2010} can go deeper in tissue ($> 700 \,\mu \rm m$). In parallel, wide-field coherent light imaging approaches of superficial microcirculation by direct recording of self-interference images, and processing of their temporal~\cite{Leutenegger2011, WangZengLiangFeng2013, LaforestDupret2013, HeNguyen2013, ChenLawLian2014} and spatial~\cite{FercherBriers1981, DunnBolay2001} fluctuations enables microvascular blood flow monitoring, deprived of depth-sectioning abilities. With a similar outcome, time-averaged heterodyne holographic interferometry in off-axis and frequency-shifting recording conditions permits frequency down-conversion of Doppler signals in the bandwidth of video cameras with high sensitivity, which enables monitoring of superficial blood flow in the cerebral cortex~\cite{AtlanGrossVitalis2006, AtlanForget2007, MagnainCastelBoucneau2014} and the retina~\cite{SimonuttiPaquesSahel2010, MagnainCastelBoucneau2014}. Nowadays, high throughput cameras provide detection bandwidths compatible with time-resolved measurements of optical phase fluctuations resulting from light-tissue interaction. In this context, and in contrast with self-interference imaging methods, off-axis holographic interferometry may circumvent the issue of imaging at extremely low irradiance levels, required to comply with exposure safety limits.

\begin{figure}[]
\centering
\includegraphics[width = 8.0 cm]{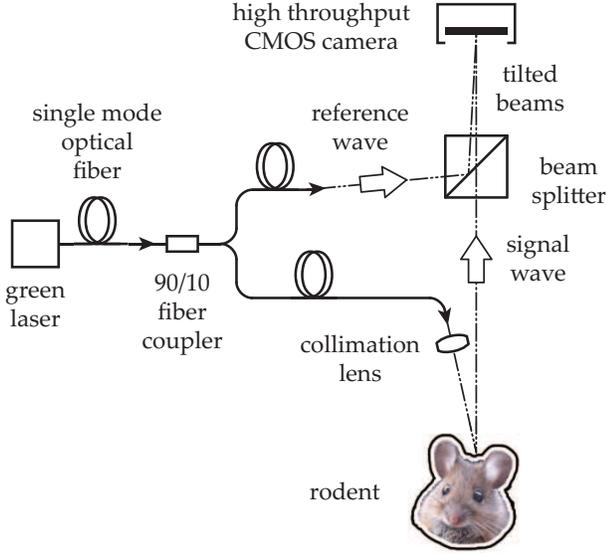}
\caption{Sketch of the fibered Mach-Zehnder holographic interferometer used for the experiment. The main laser beam is split into two channels. The optical field $E$ in the object channel, backscattered by the exposed part of the cerebral cortex of a mouse, beats against the optical field $E_{\rm LO}$ in the reference channel. Interferograms are recorded by the sensor array of a high speed camera.}
\label{fig_Setup}
\end{figure}

The experimental imaging scheme designed for this study is sketched in Fig.~\ref{fig_Setup}; it consists of a fibered Mach-Zehnder optical interferometer in off-axis configuration. The cerebral cortex of an anesthetized mouse was exposed through a $\sim$ 5 mm $\times$ 5 mm craniotomy made above the primary somatosensory cortex of the left hemisphere. The animal was anesthetized by an intraperitoneal injection of urethane ($1.7 \, \rm g/kg$). Paw withdrawal, whisker movement, and eyeblink reflexes were suppressed. A heating blanket maintained the rectally measured body temperature at 37 $^\circ$C. The skin overlying the skull was removed and the bone gently cleaned. A metal post was implanted on the occipital bone to maintain the head of the animal during the imaging session. Extreme care was taken not to damage the cerebral cortex, especially during the removal of the dura. The exposed surface of the cortex was protected by an agarose gel (1\%) and a coverslip. Experiments were performed in conformity with  the European Community Council Directive of 22nd September 2010 (010/63/UE) and approved by the local ethics committee (C2EA -59, ‘Paris Centre et Sud’, authorization number: 2012-0068). The optical source used for the experiments was a 150 mW, single-mode, fibered diode-pumped solid-state green laser (Cobolt Samba-TFB-150) at wavelength $\lambda = 532$ nm, and optical frequency $\omega_{\rm L} / (2 \pi) = 5.6 \times 10^{14} \, \rm Hz$. The preparation was illuminated with $\sim$ 3.0 mW of continuous optical power, over $\sim 4 \, {\rm mm} \times 4 \, {\rm mm}$. The reference wave is referred to as optical local oscillator (LO); its power impinging over the full sensor was $\sim 400 \, \mu {\rm W}$. The backscattered optical field $E$ was mixed with the LO field $E_{\rm LO}$ with a non-polarizing beam splitter cube, tilted by $\sim 1 ^\circ$ to ensure off-axis recording conditions. Light-tissue interaction resulted in a local phase variation $\phi(t)$ of the backscattered laser optical field $E(t) = {\cal E}  \exp \left[ i \omega_{\rm L} t + i \phi(t) \right]$, which was mixed with the LO field from the reference channel $E_{\rm LO}(t) = {\cal E}_{\rm LO} \exp \left[ i \omega_{\rm L}t \right]$. The quantities ${\cal E}$ and ${\cal E}_{\rm LO}$ are complex constants and $i$ is the imaginary unit. Optical interferograms of $768 \times 328$ pixels of coordinates $(x,y)$ were digitally acquired by the sensor array of a high throughput camera (Photron FASTCAM SA-X2, 1024 $\times$ 1024 pixels), at a frame rate of $\tau_{\rm S}^{-1} = \omega_{\rm S} / (2 \pi) = 50.0 \, \rm kHz$, with a frame exposure time of $\tau_{\rm E} = 18.98 \, \mu {\rm s}$. The distance between the preparation and the sensor was $\sim 50 \, \rm cm$. The off-axis cross-beating interferometric contribution $H = {\cal E} {\cal E}_{\rm LO}^*\exp \left[ i \phi \right]$ of the interferogram $I = \left| {\cal E} \right|^2 + \left| {\cal E}_{\rm LO} \right|^2 + H + H^*$ (where $^*$ denotes the complex conjugate) was filtered spatially~\cite{Cuche2000} from the other interferometric contributions.\\

\begin{figure}[]
\centering
\includegraphics[width = 8.0 cm]{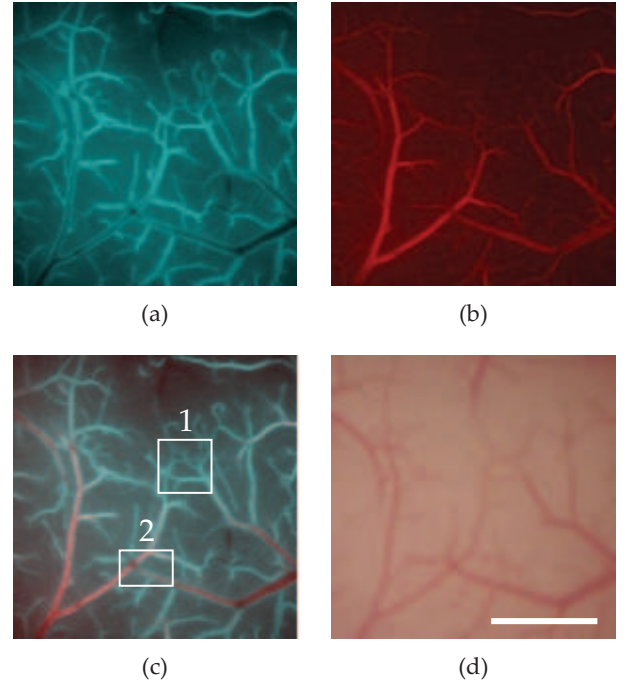}
\caption{Images of the quadratic mean of the frequency (Eq.~\ref{eq_QuadraticMeanFrequency}) calculated in the bands [2.3 kHz, 5 kHz] (a) and [5 kHz, 25 kHz] (b). Composite color Doppler image (c). White-light microscope image of the same area (d). Scale bar : 1 mm. A video of the composite Doppler image is reported in \href{https://youtu.be/9rGQ5eJGqR4}{Movie 1}.}\label{fig_CompositeFlowImage}
\end{figure}
\begin{figure}[]
\centering
\includegraphics[width = 8.0 cm]{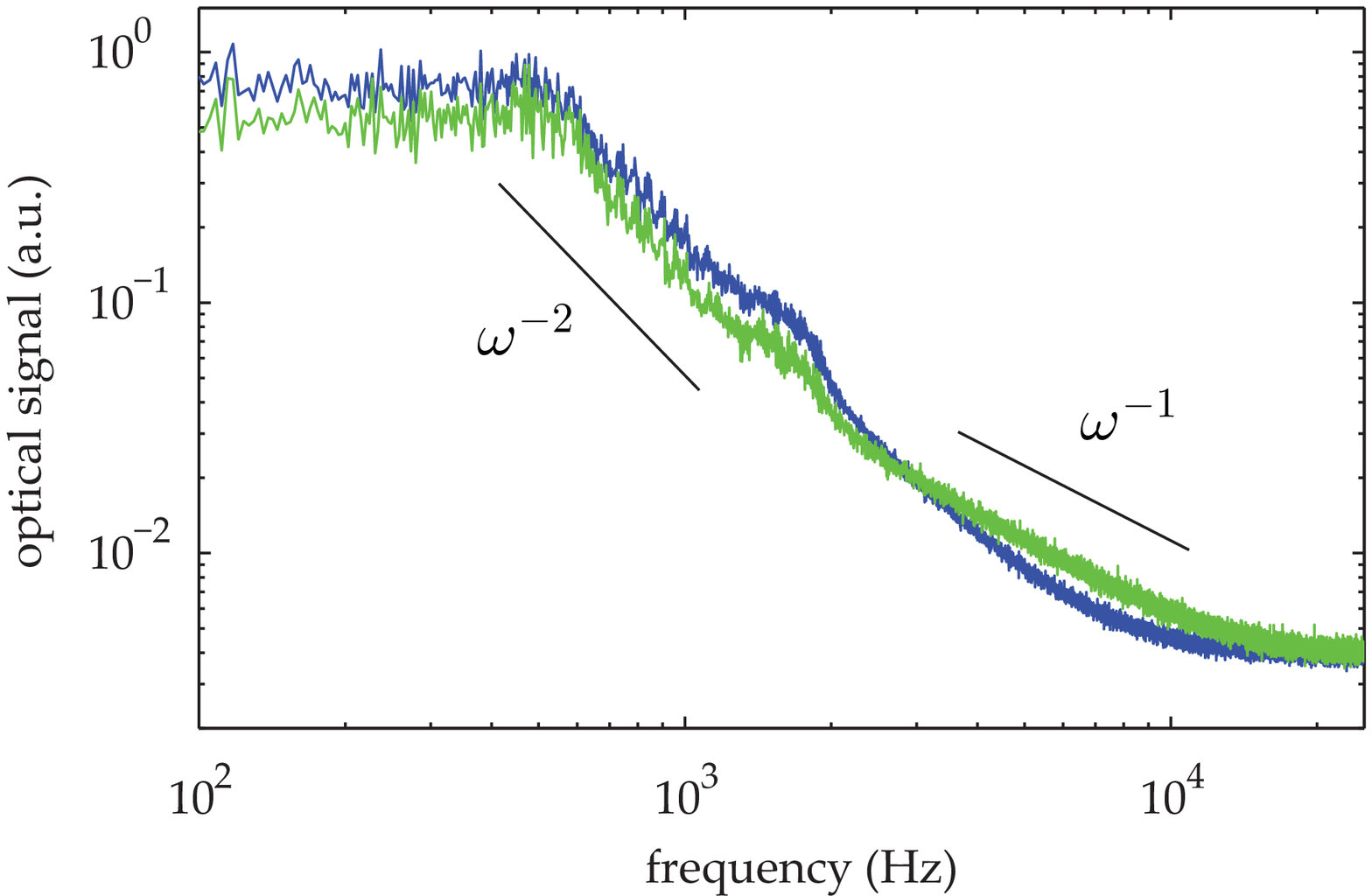}
\includegraphics[width = 8.0 cm]{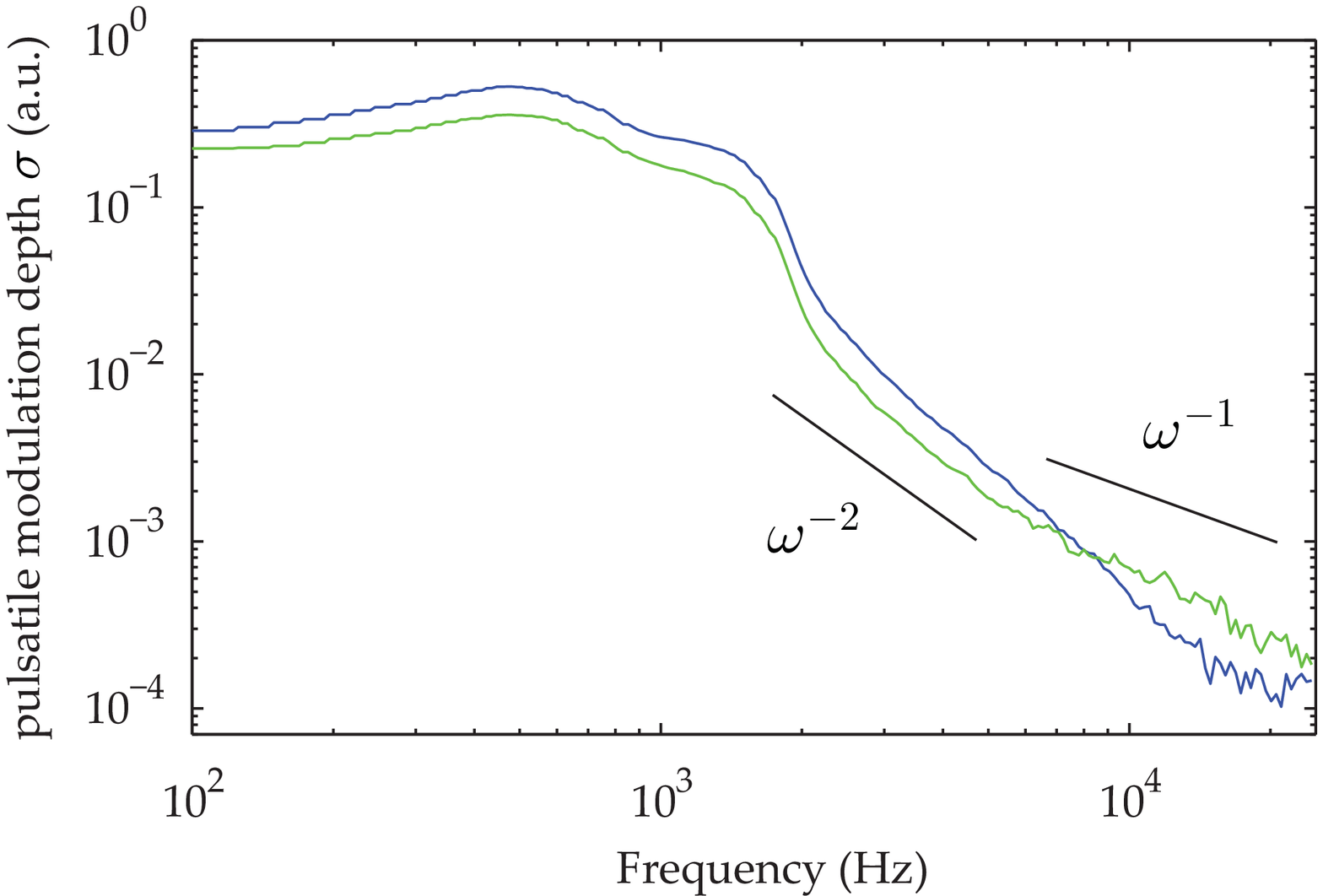}
\caption{Time-averaged optical signal $S$  (top) and pulsatile modulation depth $\sigma$ (bottom) versus frequency. Both quantities are averaged spatially in the region 1 (blue) and region 2 (green) depicted in Fig.\ref{fig_CompositeFlowImage}.}
\label{fig_Spectra}
\end{figure} 
\begin{figure*}[]
\centering
\includegraphics[width = 8.0 cm]{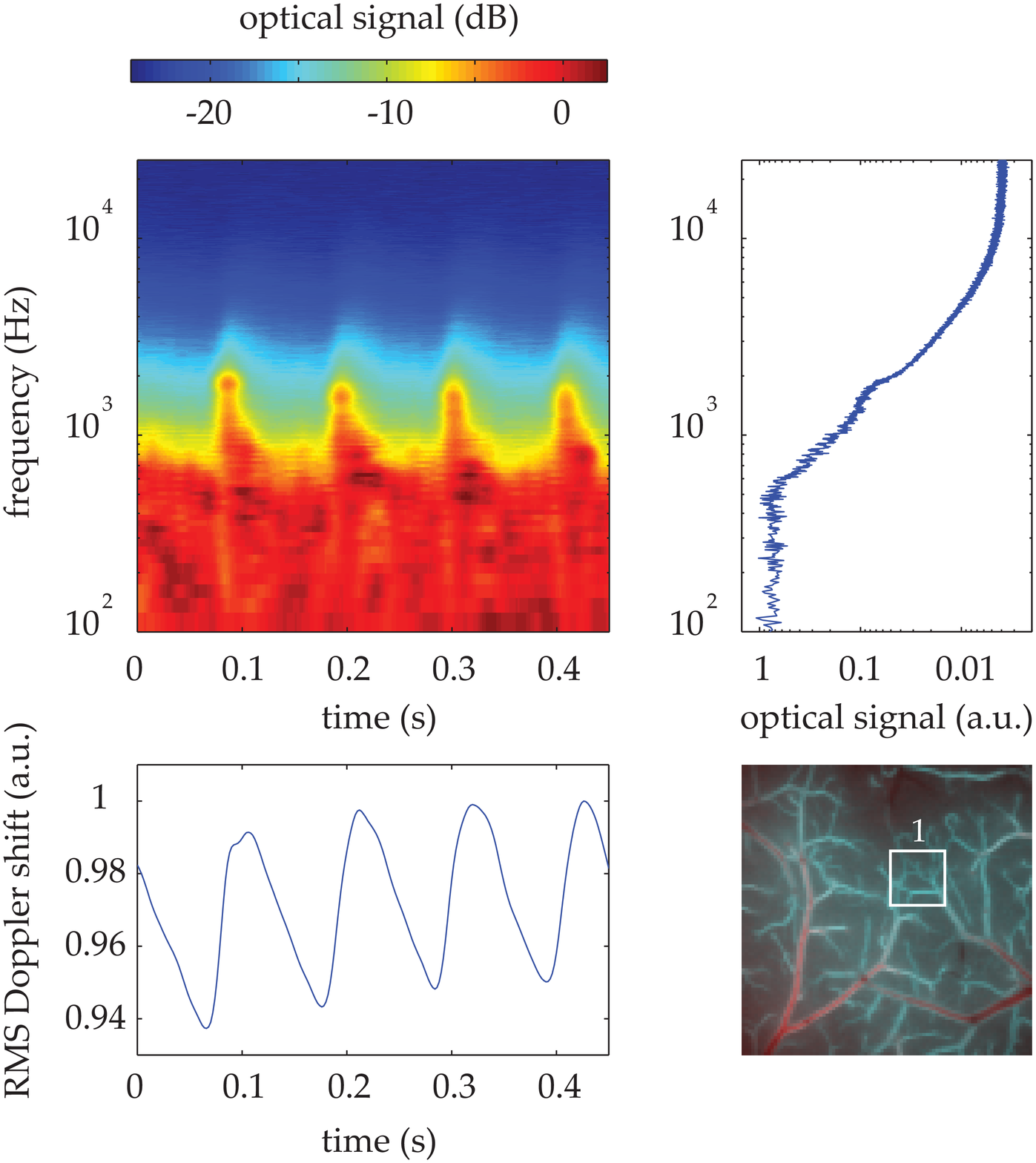}
\qquad \qquad
\includegraphics[width = 8.0 cm]{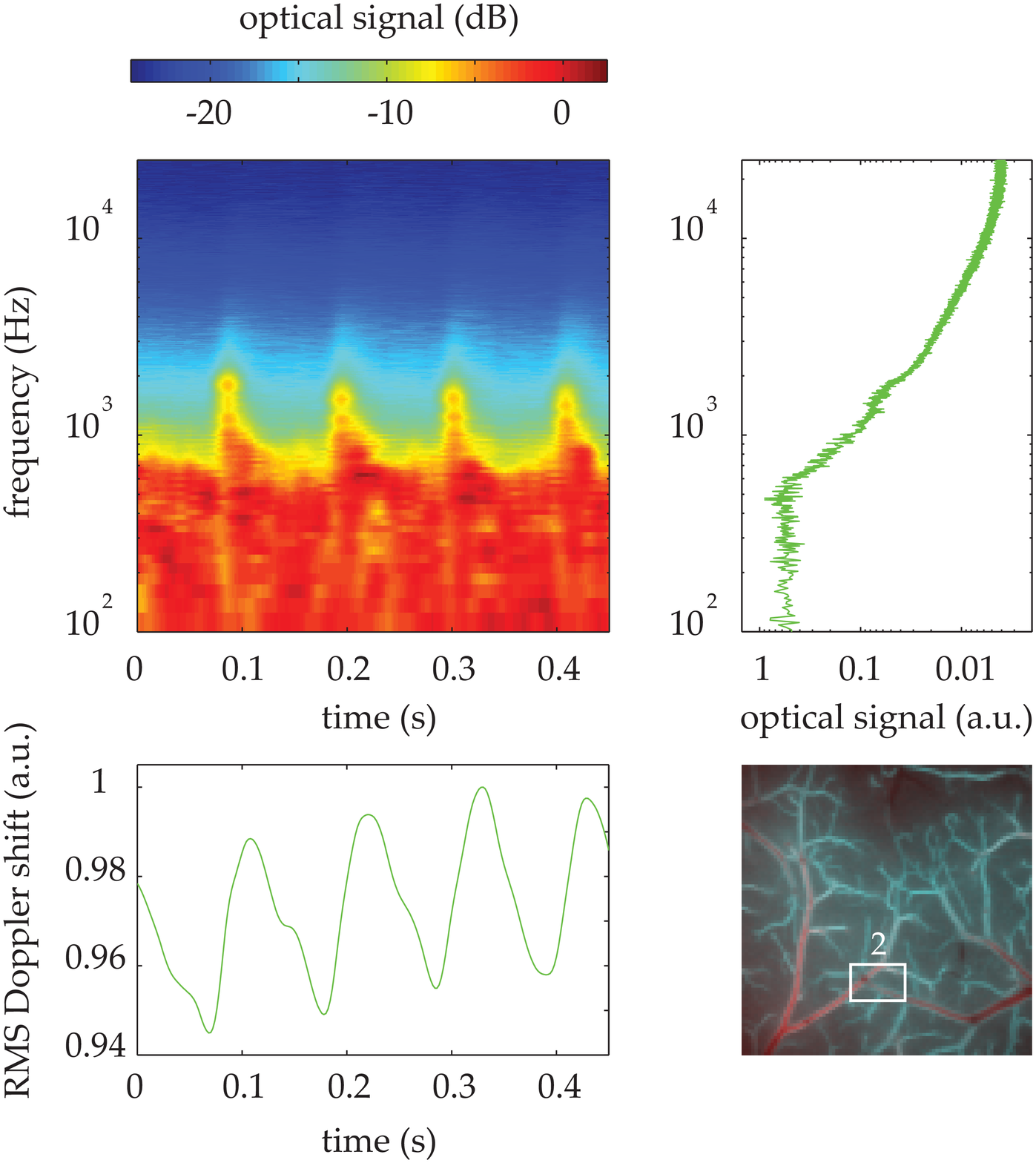}
\caption{Optical signals averaged spatially in region 1 (left pane, blue lines) and region 2 (right pane, green lines), depicted in Fig.~\ref{fig_CompositeFlowImage}. Time-frequency spectrogram $S$ (Eq.~\ref{eq_STFT}, top left), quadratic mean of the frequency $\Omega$ versus time (Eq.~\ref{eq_QuadraticMeanFrequency}, bottom left), time-averaged Fourier-transform spectrum (top right), regions of interest (bottom right).}
\label{fig_SpectrogramAndTracesROI1}
\end{figure*} 
%

 
Image rendering of off-axis holograms $H(x,y,t)$ was performed by discrete Fresnel transformation~\cite{PiedrahitaQuinteroCastanedaGarciaSucerquia2015} of recorded interferograms $I(x,y,t)$
\begin{eqnarray}\label{eq_FresnelTransform}
\nonumber H(x,y,t) = \frac{i}{\lambda z}\exp \left( -ikz \right) \iint I(x',y',t)\\
\times \exp \left[\frac{-i \pi}{\lambda z} \left((x-x')^2 + (y-y')^2\right) \right] {\rm d}x' {\rm d}y'
\end{eqnarray}
where $k = 2\pi/\lambda$ is the optical wavenumber, and $z = 0.16 \, \rm m$ is the hologram reconstruction distance. The calculation grid $(x',y')$ was zero-padded to a $1024 \times 1024$ pixels array. The squared magnitude of the short-time Fourier transform of $H$ is referred to as the optical signal
\begin{equation}\label{eq_STFT}
S(x,y,t,\omega) =\left| \int H(x,y,\tau) G(t-\tau) e^{-i \omega \tau} \, {\rm d}\tau \right|^2 
\end{equation}
where $G$ is a finite Gaussian window, centered around zero, spread over $\pm$ 2 standard deviations. The width of the frequency lineshape (assumed to be zero-mean) of the optical signal $S$ was calculated by the local quadratic mean of the frequency $\Omega = \sqrt{\left<\omega^2\right>}$, where
\begin{equation}\label{eq_QuadraticMeanFrequency}
\left<\omega^2\right> (x,y,t)  = \int S(x,y,t,\omega) \omega^2 {\rm d} \omega
\end{equation}
Assessing the perfusion from the first moments of the power spectrum of the detector signal is commonplace for laser Doppler sensors~\cite{BonnerNossal1981, Leutenegger2011}; and the Doppler signal is often normalized with the DC signal in the spectrum, in order to cancel 1- laser amplitude noise, and 2- spatial inhomogeneities of illumination of the tissue. In our case, normalization had no noticeable effect. The pulsatile modulation depth $\sigma$ of a given Doppler component $\omega$ of the signal $S$ is defined as its standard deviation in time (the square root of its variance ${\rm Var}_t$).
\begin{equation}\label{eq_PulsatileModulationDepth}
\sigma (x,y,\omega) = \left[ {\rm Var}_t \,S(x,y,t,\omega) \right]^{1/2}
\end{equation}
%


The laser Doppler signal $S$ in the image plane $(x,y)$, at time $t$, and at frequency $\omega$ (Eq.~\ref{eq_STFT}) was processed from a series of 24576 consecutive raw interferograms $I$. The temporal resolution of $\sim$ 20 ms is limited by the length of the apodization window $G$ at $\pm$ 1 standard deviation. The Shannon bandwidth of the measurement is $[ -\omega_{\rm S}/2, \omega_{\rm S}/2 [$. Images of $\Omega$ (Eq.~\ref{eq_QuadraticMeanFrequency}) are reported in Fig.~\ref{fig_CompositeFlowImage}; they were calculated in the bands [2.3 kHz, 5 kHz] (a) and [5 kHz, 25 kHz] (b). Those images are merged into the cyan and red channels of a composite color Doppler image (c), which exhibits common features with the white-light image of the same area, observed with a microscope. The temporal evolution of this image is reported in \href{https://youtu.be/9rGQ5eJGqR4}{Movie 1}. In this video, $\Omega$ was calculated with a time step of $\sim$ 2 ms, but the actual temporal resolution is limited by the width of the apodization window $\sim$ 20 ms, which also implies that temporal variations are lowpass filtered with a $\sim$ 50 Hz cutoff. In Fig.~\ref{fig_Spectra} (top), we reported the spectrum of the optical signal, averaged in the two regions of interest depicted in Fig.~\ref{fig_CompositeFlowImage}, and calculated from a time window of $\sim$ 164 ms.  In both cases, the signal is found to obey a power–law scaling $S\sim \omega^{-\beta}$, $1 \lesssim \beta \lesssim 2$. As reported previously~\cite{MagnainCastelBoucneau2014}, the Doppler spectrum is broader in regions with larger apparent vessels. A frequency sweep of local Doppler spectra of the preparation is reported in \href{https://youtu.be/Rvf6YYpYO_M}{Movie 2}. Fig.~\ref{fig_Spectra} (bottom) displays the pulsatile modulation depth $\sigma$ of the optical signal against the Doppler frequency $\omega$ (Eq.~\ref{eq_PulsatileModulationDepth}). This quantity was averaged in the two regions of interest depicted in Fig.~\ref{fig_CompositeFlowImage}(c). A frequency sweep of the local pulsatile modulation depth is reported in \href{https://youtu.be/ZLk49NCS-mc}{Movie 3}. In this video, the signal $\sigma$ was convolved with an averaging  window of $\sim$ 500 Hz, for contrast enhancement. Cardiac cycles of $\sim 0.1 \, \rm s$ appear clearly in the time-frequency spectrograms reported in Fig.~\ref{fig_SpectrogramAndTracesROI1}. The root mean square Doppler width versus time (Eq.~\ref{eq_QuadraticMeanFrequency}, calculated in the band [2.3 kHz, 25 kHz], bottom left), the time-averaged Fourier-transform spectrum (Eq.~\ref{eq_STFT}, calculated in a time window of $\sim$ 164 ms, top right), and the region of interest used for spatial averaging (bottom right) were also plotted as guides to the eye.\\


We demonstrated experimentally the measurement of superficial microvascular blood flow contrasts in the cerebral cortex of a mouse, exposed through a craniotomy. For that purpose, we designed an off-axis Mach-Zehnder holographic interferometer with a high throughput camera. This approach enabled the observation of pulsatile blood flow in small superficial blood vessels over 256 $\times$ 256 pixels with a spatial resolution of 10 microns and a temporal resolution of 20 ms. The total power of the optical radiation impinging on the preparation was 3 mW. Many experiments involving cerebral blood flow monitoring might be hindered by the need for trepanation, but this demonstration could be an important step towards the realization of a retinal blood flow imager.\\


This work was supported by Agence Nationale de la Recherche (ANR-09-JCJC-0113, ANR-11-EMMA-046), Fondation Pierre-Gilles de Gennes (FPGG014), r\'egion Ile-de-France (C’Nano, AIMA), the Investments for the Future program (LabEx WIFI: ANR-10-LABX-24, ANR-10-IDEX-0001-02 PSL*), and European Research Council (ERC Synergy HELMHOLTZ).


\begin{thebibliography}{10}

\bibitem{OgawaLee1990}
Seiji Ogawa, Tso-Ming Lee, Alan~R Kay, and David~W Tank.
\newblock Brain magnetic resonance imaging with contrast dependent on blood
  oxygenation.
\newblock {\em Proceedings of the National Academy of Sciences},
  87(24):9868--9872, 1990.

\bibitem{MaceMontaldoCohen2011}
Emilie Mac{\'e}, Gabriel Montaldo, Ivan Cohen, Michel Baulac, Mathias Fink, and
  Mickael Tanter.
\newblock Functional ultrasound imaging of the brain.
\newblock {\em Nature Methods}, 8(8):662--664, 2011.

\bibitem{MaceMontaldoOsmanski2013}
Emilie Mace, Gabriel Montaldo, B~Osmanski, Israel Cohen, Mathias Fink, and
  Mickael Tanter.
\newblock Functional ultrasound imaging of the brain: theory and basic
  principles.
\newblock {\em Ultrasonics, Ferroelectrics, and Frequency Control, IEEE
  Transactions on}, 60(3):492--506, 2013.

\bibitem{Wang2010}
Ruikang~K Wang.
\newblock Optical microangiography: a label-free 3-d imaging technology to
  visualize and quantify blood circulations within tissue beds in vivo.
\newblock {\em Selected Topics in Quantum Electronics, IEEE Journal of},
  16(3):545--554, 2010.

\bibitem{DziennisQinShi2015}
Suzan Dziennis, Jia Qin, Lei Shi, and Ruikang~K Wang.
\newblock Macro-to-micro cortical vascular imaging underlies regional
  differences in ischemic brain.
\newblock {\em Scientific Reports}, 5, 2015.

\bibitem{MaslovZhang2008}
Konstantin Maslov, Hao~F Zhang, Song Hu, and Lihong~V Wang.
\newblock Optical-resolution photoacoustic microscopy for in vivo imaging of
  single capillaries.
\newblock {\em Optics letters}, 33(9):929--931, 2008.

\bibitem{YaoMaslov2010}
Junjie Yao, Konstantin~I Maslov, Yunfei Shi, Larry~A Taber, and Lihong~V Wang.
\newblock In vivo photoacoustic imaging of transverse blood flow by using
  doppler broadening of bandwidth.
\newblock {\em Optics letters}, 35(9):1419--1421, 2010.

\bibitem{Leutenegger2011}
M.~Leutenegger, E.~Martin-Williams, P.~Harbi, T.~Thacher, W.~Raffoul,
  M.~Andr{\'e}, A.~Lopez, P.~Lasser, and T.~Lasser.
\newblock Real-time full field laser doppler imaging.
\newblock {\em Biomedical Optics Express}, 2(6):1470--1477, 2011.

\bibitem{WangZengLiangFeng2013}
Mingyi Wang, Yaguang Zeng, Xianjun Liang, Guanping Feng, Xuanlong Lu, Junbo
  Chen, Dingan Han, and Guojian Yang.
\newblock In vivo label-free microangiography by laser speckle imaging with
  intensity fluctuation modulation.
\newblock {\em Journal of Biomedical Optics}, 18(12):126001--126001, 2013.

\bibitem{LaforestDupret2013}
Timothe Laforest, Antoine Dupret, Arnaud Verdant, Fran{\c{c}}ois Ramaz, Sylvain
  Gigan, Gilles Tessier, and Emilie~Benoit la~Guillaume.
\newblock A 4000 hz cmos image sensor with in-pixel processing for light
  measurement and modulation.
\newblock pages 1--4, 2013.

\bibitem{HeNguyen2013}
Diwei He, Hoang~C Nguyen, Barrie~R Hayes-Gill, Yiqun Zhu, John~A Crowe, Cally
  Gill, Geraldine~F Clough, and Stephen~P Morgan.
\newblock Laser doppler blood flow imaging using a cmos imaging sensor with
  on-chip signal processing.
\newblock {\em Sensors}, 13(9):12632--12647, 2013.

\bibitem{ChenLawLian2014}
DG~Chen, MK~Law, Y~Lian, and A~Bermak.
\newblock Low-power cmos laser doppler imaging using non-cds pixel readout and
  13.6-bit sar adc.
\newblock {\em IEEE transactions on biomedical circuits and systems}, 2014.

\bibitem{FercherBriers1981}
A.~F. Fercher and J.~D. Briers.
\newblock Flow visualisation by means of single-exposure speckle photography.
\newblock {\em Opt. Commun.}, 37:326--330, 1981.

\bibitem{DunnBolay2001}
A.~Dunn, H.~Bolay, M.~A. Moskowitz, and D.~A. Boas.
\newblock Dynamic imaging of cerebral blood flow using laser speckle.
\newblock {\em Jounal of Cerebral Blood Flow and Metabolism}, 21(3):195--201,
  2001.

\bibitem{AtlanGrossVitalis2006}
M.~Atlan, M.~Gross, T.~Vitalis, A.~Rancillac, B.~C. Forget, and A.~K. Dunn.
\newblock Frequency-domain, wide-field laser doppler in vivo imaging.
\newblock {\em Optics Letters}, 31(18):2762--2764, 2006.

\bibitem{AtlanForget2007}
Michael Atlan, Benoit~C. Forget, Albert~C. Boccara, Tania Vitalis, Armelle
  Rancillac, Andrew~K. Dunn, and Michel Gross.
\newblock Cortical blood flow assessment with frequency-domain laser doppler
  microscopy.
\newblock {\em Journal of Biomedical Optics}, 12(2):024019, 2007.

\bibitem{MagnainCastelBoucneau2014}
Caroline Magnain, Amandine Castel, Tanguy Boucneau, Manuel Simonutti, Isabelle
  Ferezou, Armelle Rancillac, Tania Vitalis, Jos{\'e}-Alain Sahel, Michel
  Paques, and Michael Atlan.
\newblock Holographic laser doppler imaging of microvascular blood flow.
\newblock {\em JOSA A}, 31(12):2723--2735, 2014.

\bibitem{SimonuttiPaquesSahel2010}
M.~Simonutti, M.~Paques, J.~A. Sahel, M.~Gross, B.~Samson, C.~Magnain, and
  M.~Atlan.
\newblock Holographic laser doppler ophthalmoscopy.
\newblock {\em Opt. Lett.}, 35(12):1941--1943, 2010.

\bibitem{Cuche2000}
Etienne Cuche, Pierre Marquet, and Christian Depeursinge.
\newblock Spatial filtering for zero-order and twin-image elimination in
  digital off-axis holography.
\newblock {\em Applied Optics}, 39(23):4070, 2000.

\bibitem{PiedrahitaQuinteroCastanedaGarciaSucerquia2015}
Pablo Piedrahita-Quintero, Raul~Casta\ {n}eda, and Jorge Garcia-Sucerquia.
\newblock Numerical wave propagation in imagej.
\newblock {\em Appl. Opt.}, 54(21):6410--6415, Jul 2015.

\bibitem{BonnerNossal1981}
R.~Bonner and R.~Nossal.
\newblock Model for laser doppler measurements of blood flow in tissue.
\newblock {\em Applied Optics}, 20:2097--2107, 1981.

\end{thebibliography}

\end{document}